\documentclass[pre,superscriptaddress,twocolumn,amsmath,amssymb,floatfix]{revtex4}

\usepackage{graphicx}
\usepackage{color}

\begin{document}

\title{Correlated bursts and the role of memory range} 
\author{Hang-Hyun Jo}
\affiliation{BK21plus Physics Division and Department of Physics, Pohang University of Science and Technology, Pohang 790-784, Republic of Korea}
\affiliation{Department of Computer Science, Aalto University School of Science, P.O. Box 15500, Espoo, Finland}
\author{Juan I. Perotti}
\affiliation{Department of Computer Science, Aalto University School of Science, P.O. Box 15500, Espoo, Finland}
\affiliation{IMT Institute for Advanced Studies Lucca, Piazza San Francesco 19, I-55100, Lucca, Italy}
\author{Kimmo Kaski}
\affiliation{Department of Computer Science, Aalto University School of Science, P.O. Box 15500, Espoo, Finland}
\author{J\'anos Kert\'esz}
\affiliation{Center for Network Science, Central European University, N\'ador u. 9, H-1051 Budapest, Hungary}
\affiliation{Department of Computer Science, Aalto University School of Science, P.O. Box 15500, Espoo, Finland}

\date{\today}

\begin{abstract}
Inhomogeneous temporal processes in natural and social phenomena have been described by bursts that are rapidly occurring events within short time-periods alternating with long periods of low activity. In addition to the analysis of heavy-tailed interevent time distributions, higher-order correlations between interevent times, called \emph{correlated bursts}, have been studied only recently. As the underlying mechanism behind such correlated bursts is far from being fully understood, we devise a simple model for correlated bursts using a self-exciting point process with variable range of memory. Whether a new event occurs is stochastically determined by a memory function that is the sum of decaying memories of past events. In order to incorporate the noise and/or limited memory capacity of systems, we apply two memory loss mechanisms: fixed number or variable number of memories. By analysis and numerical simulations, we find that too much memory effect may lead to a Poissonian process, implying that there exists an intermediate range of memory effect to generate correlated bursts comparable to empirical findings. Our conclusions provide deeper understanding of how long-range memory affects correlated bursts.
\end{abstract}


\maketitle

\section{Introduction}

Many natural phenomena and human activities are extremely inhomogeneous in time. Solar flares, earthquakes~\cite{deArcangelis2006}, firing of neurons~\cite{Kemuriyama2010}, and human communication~\cite{Barabasi2005} are just some examples. In all these phenomena events occurring within short time periods, called bursts, are alternating with random, long periods of low activity~\cite{Barabasi2005}. Often the elements behaving in this manner constitute a temporal network~\cite{Holme2012} and then
the processes such as spreading on the network are strongly influenced by the burstiness of the time series~\cite{Karsai2011,Miritello2011,Starnini2012,Jo2014,Moinet2015}. Or burstiness can be influenced by spreading~\cite{Garcia-Perez2015}. The natural question raises: How to characterize the highly inhomogeneous dynamics and how to model it? This is important for discovering analogies between different processes leading to possible universalities and for understanding the effect of temporal inhomogeneities on the network processes.

At the simplest level, the bursty time series is characterized by the heavy-tailed interevent time distribution $P(\tau)$, where $\tau$ is the time interval between two consecutive events. $P(\tau)$ has often been described by a power law:
\begin{equation}
P(\tau)\sim \tau^{-\alpha}.
\end{equation}
However, the interevent time distribution does not provide a complete characterization. The higher order description of bursts focuses on dependencies between interevent times, i.e., higher order memory effects. These can be approached in different ways. One possibility is to calculate the autocorrelation function. For this approach Goh and Barab\'asi defined a memory coefficient measuring short-range memory effect~\cite{Goh2008} as following:
\begin{equation}
  M = \frac{\langle (\tau_i-\langle\tau\rangle)(\tau_{i+1}-\langle\tau\rangle)\rangle}{\sigma^2}, 
  \label{eq:memoryCoeff}
\end{equation}
where $\tau_i$ denotes the $i$th interevent time, and $\langle\tau\rangle$ and $\sigma$ are the average and standard deviation of interevent times. The aim was to characterize the bursty time series by two quantities, $M$ and the burstiness parameter $B$, defined as
\begin{equation}
B = \frac{\sigma - \langle\tau\rangle}{\sigma + \langle\tau\rangle}.
\label{eq:burstiness}
\end{equation}
It was found in many human activities that $M$ is close to $0$. To describe long-range memory effects, one can use the entire autocorrelation function of the time series. Recently, it was shown that the power-law decay of the autocorrelation function is implied by a power-law interevent time distribution even without any correlations between consecutive interevent times. Precisely, the scaling relation $\alpha+\gamma=2$ was proven with $\gamma$ denoting the decaying exponent of the autocorrelation function~\cite{Vajna2013}. However, more works need to be done for the validity of the scaling relation, as the effect of dependencies between interevent times in the bursty time series is not yet fully understood.

An approach sensitive to dependencies was recently introduced by using the notion of bursty trains~\cite{Karsai2012}. A bursty train is defined as a set of events, such that any pair of consecutive events in the train is separated by an interevent time within a given time window $\Delta t$. The distribution of the number $E$ of events in the trains follows an exponential function if the interevent times are independently and identically distributed. It was found, however, that in many empirical cases $E$ is power-law distributed, i.e.,
\begin{equation}
  P_{\Delta t}(E)\sim E^{-\beta} 
\end{equation}
for a wide range of $\Delta t$. This notion of \emph{correlated bursts} was empirically observed in earthquakes, neuronal activities, and human communication patterns~\cite{Karsai2012}. Such correlations are clearly due to memory effects. 

Generative models for correlated bursts have been devised and studied. Karsai \emph{et al.} introduced two-state model with memory function~\cite{Karsai2012}: One state is for generating power-law distributed interevent times that are uncorrelated, while the other state is for generating short-timescale bursty trains. For the latter, they define a memory function as the probability of generating one more event in the train provided that $l$ events have already been generated in the train: 
\begin{equation}
  \label{eq:p_n}
  p(l)=\left(\frac{l}{l+1}\right)^\nu.
\end{equation}
Here $\alpha$ and $\beta$ are directly controlled by parameters for memory functions, e.g., $\beta=\nu+1$. In this model, the onset of bursty trains is assumed to be known or at least declared in order to use the above memory function, requiring additional information. Such assumption is not necessarily the case in reality. Thus in this paper we suggest more natural and intuitive mechanism for correlated bursts that does not need declaring bursty trains. We also investigate the robustness of the scaling relation $\alpha+\gamma=2$ with respect to the strength of dependencies between interevent times.

\section{Model}

We study a generative model for correlated bursts with variable range of memory effect. In our model, bursty trains emerge from the stochastic process using a memory function. Note that our memory function has nothing to do with Eq.~(\ref{eq:p_n}). At time step $t=1$, the first event occurs, and the $i$th event occurs at time step $t=t_i$. The probability that the $(n+1)$th event occurs at time step $t$ is given by
\begin{eqnarray}
  \label{eq:pmt}
  p[m(t)]&=&1-e^{-\mu m(t)-\epsilon}\\
  \label{eq:memory}
  m(t)&=&\sum_{i=1}^n \frac{1}{t-t_i}\ \textrm{for}\ t>t_n,
\end{eqnarray}
where $m(t)$ denotes the \emph{memory function} that is the time-weighted sum of all the past events. Accordingly, $\mu$ controls the strength of memory effect, such that the larger $\mu$ implies the stronger memory effect. Here we use $\epsilon=10^{-6}$ to indicate spontaneous events taking place with very small probability. Once the $(n+1)$th event occurs, the memory term due to this event is added to the memory function. Note that $t$ is discrete and $t-t_i\geq 1$.

We remark that our model can be considered as a self-exciting point process~\cite{Adamopoulos1976,Utsu1995,Ogata2006} with a power-law kernel. These processes have been extensively studied for earthquakes~\cite{Shcherbakov2005,Saichev2006,Touati2009,Bottiglieri2010,Lippiello2012} as well as in application to social systems~\cite{Crane2008,Masuda2012}. In such processes, the time-varying event rate is given as a function of the past events. As for the kernel, the Omori's law
has been widely used, stating that aftershock frequency decreases with an elapsed time after the main shock, e.g., in a form of $(t-t_i)^{-1-\theta}$ with small positive $\theta$~\cite{Saichev2006}. The self-exciting point process with Omori's law is called epidemic-type aftershock sequence (ETAS) model. Note that our memory function in Eq.~(\ref{eq:memory}) corresponds to the case with $\theta=0$. Analytic and numerical approaches to the ETAS model have shown that interevent time distributions
are mostly described by a Gamma function~\cite{Lippiello2012}, implying that $\alpha\leq 1$. However, one finds evidence for $\alpha>1$ in many other natural and social phenomena~\cite{Oliveira2005,Dezso2006,Zhou2008}. Despite the similarity between our model and the ETAS model, our model shows different results such as $\alpha>1$.

As a novel feature compared to the family of ETAS models, we introduce memory loss mechanisms as most systems may lose their memory by various reasons, e.g., noise, limited memory capacities, or periodic resetting in circadian cycles of humans~\cite{Jo2012}. 
We incorporate the \emph{sequential memory loss} mechanism by considering only a finite number $L$ of terms in Eq.~(\ref{eq:memory}):
\begin{equation}
  m(t)=\sum_{i=n-L+1}^n \frac{1}{t-t_i}\ \textrm{for}\ t>t_n.
\end{equation}
This implies that once an event occurs, the memory due to the oldest event is immediately lost. Here $L=1$ implies no memory before the latest event. Note that without memory loss, i.e., when $L$ is infinite, $m(t)$ might diverge as $\ln t$. 

We can consider more realistic memory loss mechanisms depending on the systems of interest. For example, the condition of the fixed $L$ can be relaxed by considering variable $L$. Whenever an event occurs, $m(t)$ is initialized except for the latest event, i.e., by setting $L=1$, with a probability
\begin{equation}
  \label{eq:q_L}
  q[L(t)]=1-\left[\frac{L(t)}{L(t)+1}\right]^\nu +\epsilon_L,
\end{equation}
where $L(t)$ is the number of terms in memory function at time $t$. This can be called \emph{preferential memory loss} mechanism. Here we have introduced the spontaneous initialization of $m(t)$ with very small $\epsilon_L=10^{-6}$, otherwise if $\epsilon_L=0$, $q(L)$ may be extremely small due to extremely large $L(t)$ and vice versa. With this $q(L)$, we expect that the distribution of $L$ is proportional to $L^{-\nu-1}e^{-L/L_c}$ with $L_c\equiv \epsilon_L^{-1}$. We will study both memory loss mechanisms one by one.

We remark that our model is intrinsically non-stationary due to the long-range memory effect. However, non-stationary periods are limited to timescales of the order of $\epsilon^{-1}$, as to be numerically confirmed by the decaying behavior of autocorrelation function for the delay time of the order of $\epsilon^{-1}$.

\section{Results}

\subsection{Sequential memory loss}

\begin{figure}[!t]
  \includegraphics[width=.9\columnwidth]{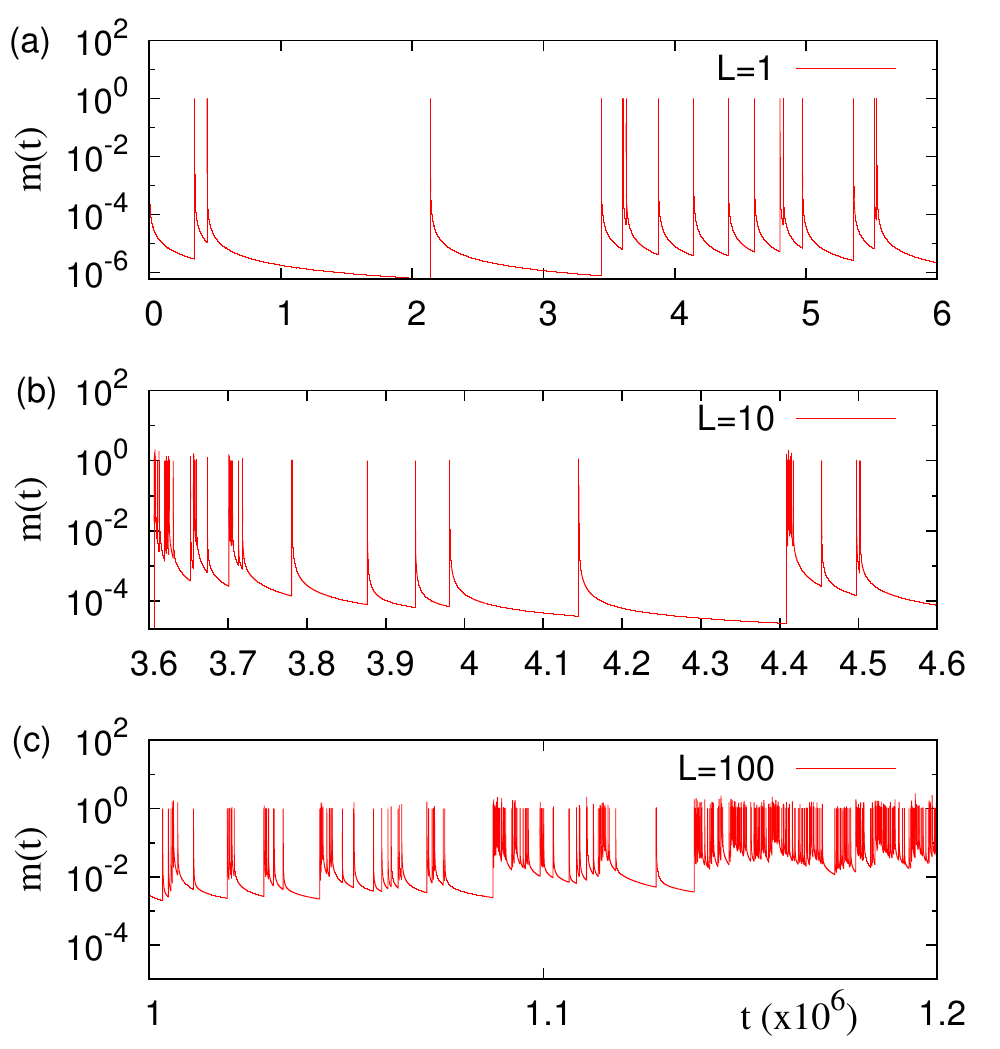}
  \caption{(Color online) Temporal evolution of memory function $m(t)$ in the model with sequential memory loss mechanism, where the number of memories is denoted by $L$. We used $\mu=1/10$, and $L=1$ (a), $10$ (b), and $100$ (c).}
    \label{fig:sequentialML_memory}
\end{figure}

\begin{figure*}[!t]
    \includegraphics[width=.75\textwidth]{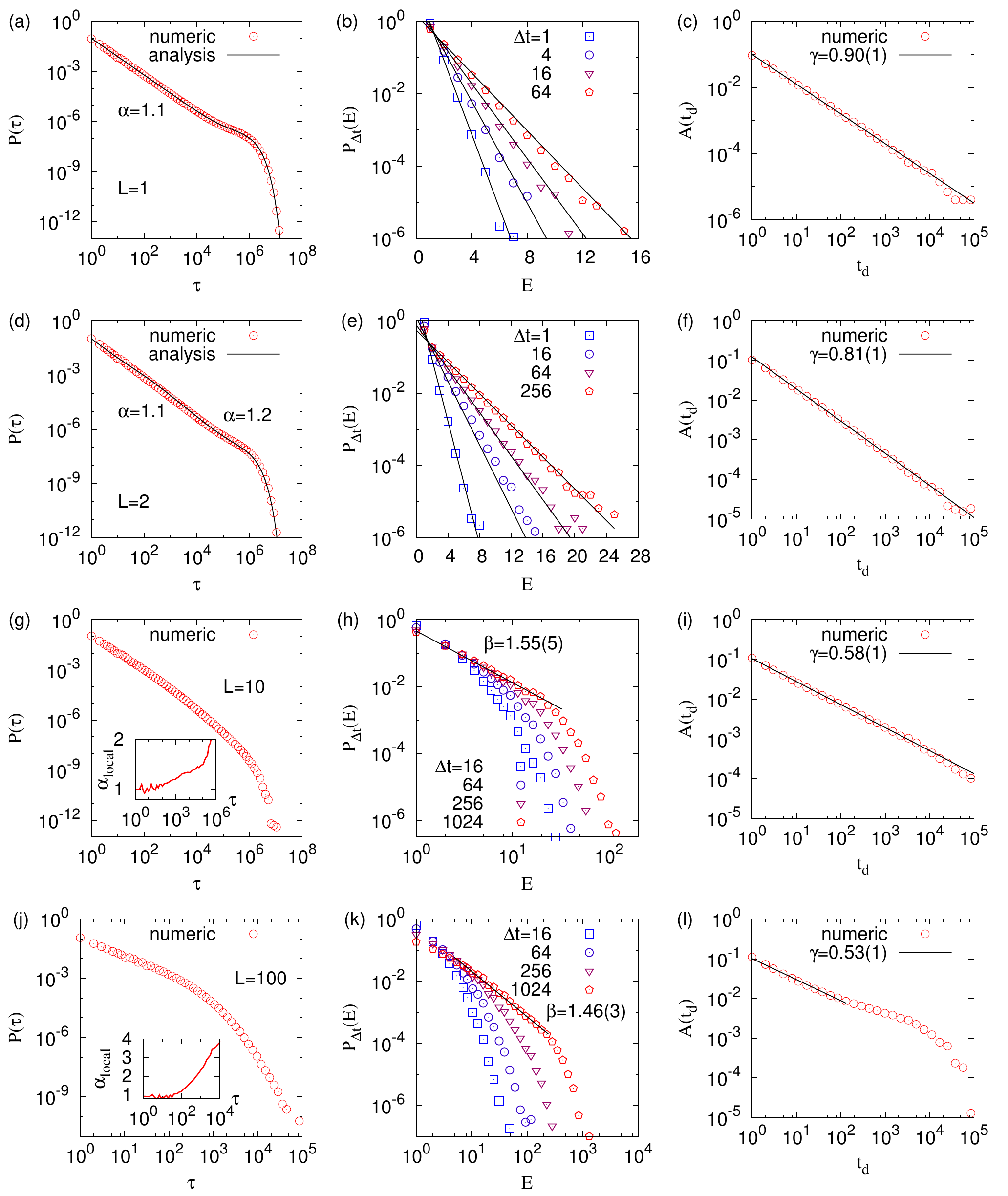}
  \caption{(Color online) Interevent time distributions $P(\tau)$ (left), bursty train distributions $P_{\Delta t}(E)$ (center), and autocorrelation functions $A(t_d)$ (right) in the model with sequential memory loss mechanism, where the number of memories is denoted by $L$. We used $\mu=1/10$, and $L=1$, $2$, $10$, and $100$ (from top to bottom).}
  \label{fig:sequentialML_mu0_1}
\end{figure*}

In general, the probability of finding an interevent time $\tau$ between events occurred at $t_n$ and $t_n+\tau$ is written as
\begin{equation}
  \label{eq:pt_tau}
  P(\tau)=\left[\prod_{t=t_n+1}^{t_n+\tau-1}e^{-\mu m(t)-\epsilon}\right] \left[1-e^{-\mu m(t_n+\tau)-\epsilon}\right].
\end{equation}
This formula is exact as the model is defined in the discrete time $t$, while the formula for continuous time can be found in~\cite{Shcherbakov2005}. Let us first consider the simplest case when the model has no memory before the latest event, i.e., $L=1$. Since the distribution does not depend on $t_n$ but only on $t-t_n$, we set $t_n=0$ without loss of generality. Then, we use 
\begin{equation}
  m(t)=\frac{1}{t}. 
\end{equation}
The numerical result of $m(t)$ is depicted in Fig.~\ref{fig:sequentialML_memory}(a). This can be related to time-varying priority queuing models studied in~\cite{Jo2012c}, where the decaying priority of the task was considered as $\sim t^{-a}$. One gets the interevent time distribution:
\begin{eqnarray}
  \nonumber
  P(\tau)&=&\left[\prod_{t=1}^{\tau-1}e^{-\mu/t-\epsilon}\right] \left(1-e^{-\mu/\tau-\epsilon}\right)\\
  \nonumber
  &\approx&\exp\left[-\mu\ln(\tau-1) -\epsilon(\tau-1)\right] \left(\frac{\mu}{\tau}+\epsilon\right)\\
  &\approx&\left[\mu\tau^{-(1+\mu)} + \frac{\tau^{-\mu}}{\tau_c}\right]e^{-\tau/\tau_c},\ \tau_c\equiv\epsilon^{-1}.
\end{eqnarray}
In the last line, we have assumed $\tau\gg 1$. This analytic result perfectly fits the numerical results even for small values of $\tau$, see Fig.~\ref{fig:sequentialML_mu0_1}(a). For numerical simulations, we have generated the event sequence consisting of up to $10^6$ events using $\mu=1/10$ for all cases. The bump observed for large $\tau$ is clearly due to the Poisson events with positive $\epsilon$. We find the power-law exponent of interevent time distribution to be
\begin{equation}
  \label{eq:alphaL1}
  \alpha=\left\{\begin{tabular}{ll}
    $1+\mu$ & for $\tau\ll\mu\tau_c$,\\
    $\mu$ & for $\mu\tau_c\ll\tau\ll\tau_c$.
  \end{tabular}\right.
\end{equation}
When $\tau_c=10^6$ and $\mu=1/10$, the scaling regime with $\alpha=\mu$ turns out to be almost invisible. Thus the dominant scaling behavior is characterized by $\alpha=1+\mu$. 

Since all interevent times are fully uncorrelated, the bursty train distribution is given by~\cite{Karsai2012}
\begin{eqnarray}
  P_{\Delta t}(E)&=&F(\Delta t)^{E-1}\left[1-F(\Delta t)\right],\\
  F(\Delta t)&\equiv &\sum_{\tau=1}^{\Delta t}P(\tau).
\end{eqnarray}
For $E\gg 1$, one gets the exponential distribution of bursty trains as
\begin{equation}
  P_{\Delta t}(E)\approx e^{-E/E_c(\Delta t)} \left[1-e^{-1/E_c(\Delta t)}\right]
\end{equation}
with $E_c(\Delta t)\equiv -[\ln F(\Delta t)]^{-1}$, which is numerically confirmed in Fig.~\ref{fig:sequentialML_mu0_1}(b). In case of $\epsilon=0$, we have $E_c(\Delta t)\approx (\Delta t)^{\mu}$ for $\Delta t\gg 1$.

In order to calculate the autocorrelation function, we first denote the event sequence by $x(t)$ that has the value of $1$ at the moment of event occurred, $0$ otherwise. The autocorrelation function with delay time $t_d$ is defined as
\begin{equation}
  A(t_d)=\frac{ \langle x(t)x(t+t_d)\rangle_t- \langle x(t)\rangle^2_t}{ \langle x(t)^2\rangle_t- \langle x(t)\rangle^2_t}.
\end{equation}
For the power-law interevent time distribution, one may find that $A(t_d)\sim t_d^{-\gamma}$. For the uncorrelated interevent times, it has been proved that $\alpha+\gamma=2$~\cite{Vajna2013}. This scaling relation is numerically confirmed with the estimated value of $\gamma$ in Fig.~\ref{fig:sequentialML_mu0_1}(c).

Next, we consider the case of $L=2$, when the memory function is composed of two terms corresponding to the latest event and the second latest event, respectively. The interevent time between those two events is denoted by $\tau_1$. Again we set $t_n=0$ in Eq.~(\ref{eq:pt_tau}). The \emph{conditional} memory function reads
\begin{equation}
  m(t|\tau_1)= \frac{1}{t}+ \frac{1}{t+\tau_1},
\end{equation}
leading to the \emph{conditional} interevent time distribution $P(\tau|\tau_1)$ as
\begin{eqnarray}
  P(\tau|\tau_1)&\approx& e^{-\mu f(\tau|\tau_1)-\tau/\tau_c}[\mu g(\tau|\tau_1)+\tau_c^{-1}],\label{eq:Ptautau1}\\
  f(\tau|\tau_1)&\equiv &\ln \tau+\ln (\tau+\tau_1)-\ln \tau_1,\\
  g(\tau|\tau_1)&\equiv &\frac{1}{\tau}+\frac{1}{\tau+\tau_1}.
\end{eqnarray}
If $\tau\ll\tau_1$, we get
\begin{equation}
  \label{eq:Ptautau2}
  P(\tau|\tau_1)\approx \mu\tau^{-(1+\mu)}+\left(\frac{\mu}{\tau_1}+\frac{1}{\tau_c}\right)\tau^{-\mu}.
\end{equation}
On the other hand, if $\tau\gg\tau_1$, we get
\begin{equation}
  \label{eq:Ptautau3}
  P(\tau|\tau_1)\approx \tau_1^{\mu} \left[2\mu\tau^{-(1+2\mu)}+\frac{\tau^{-2\mu}}{\tau_c}\right] e^{-\tau/\tau_c}.
\end{equation}
Then, $P(\tau)$ could be obtained by solving the following self-consistent equation:
\begin{equation}
  P(\tau)=\sum_{\tau_1} P(\tau|\tau_1)P(\tau_1),
\end{equation}
which is however not trivial. Instead we find that the leading term of Eq.~(\ref{eq:Ptautau2}) is not explicitly dependent on $\tau_1$, and that $\tau_1^{\mu}$ appears in Eq.~(\ref{eq:Ptautau3}) only as a coefficient. Thus we expect that $P(\tau)\approx P(\tau|\tau_\times)$ with $\tau_1$ in Eq.~(\ref{eq:Ptautau1}) replaced by a crossover interevent time $\tau_\times$. We numerically estimate $\tau_\times\approx 70.2$ by fitting $P(\tau|\tau_\times)$ to the simulation result of $P(\tau)$, shown in Fig.~\ref{fig:sequentialML_mu0_1}(d). In sum, provided that $\tau_\times \ll 2\mu\tau_c\ll \tau_c$, one can get the power-law exponent:
\begin{equation}
  \alpha=\left\{\begin{tabular}{ll}
    $1+\mu$ & for $\tau\ll\tau_\times$,\\
    $1+2\mu$ & for $\tau_\times\ll\tau\ll 2\mu\tau_c$,\\
    $2\mu$ & for $2\mu\tau_c\ll\tau\ll\tau_c$.
  \end{tabular}\right.
  \label{eq:alphaL2}
\end{equation}

\begin{figure}[!t]
  \includegraphics[width=\columnwidth]{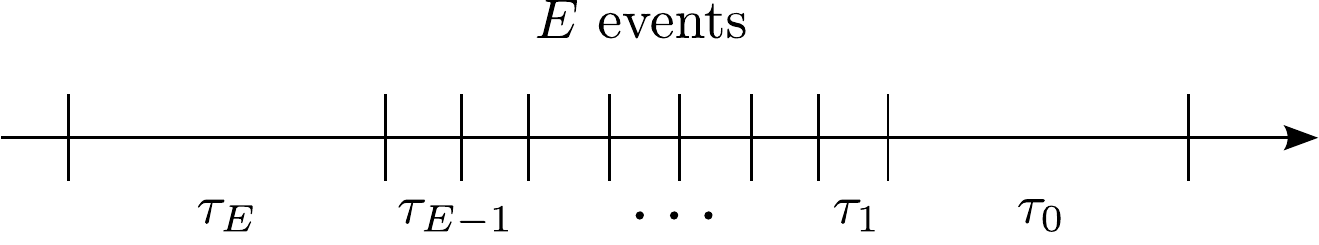}
  \caption{An event train of $E$ events, where $\tau_0,\ \tau_E>\Delta t$ and $\tau_i\leq \Delta t$ for $i=1,\cdots,E-1$.}
    \label{fig:event_train}
\end{figure}

The bursty train distribution can be calculated as
\begin{equation}
  P_{\Delta t}(E)=C\sum_{\tau_0,\tau_E=\Delta t+1}^\infty \sum_{\tau_1,\cdots,\tau_{E-1}=1}^{\Delta t} P(\tau_E) \prod_{i=0}^{E-1}P(\tau_{i}|\tau_{i+1})
  \label{eq:PE_L2}
\end{equation}
with a normalization constant $C$. An example of event train is shown in Fig.~\ref{fig:event_train}. Here we decompose the interevent times in Eq.~(\ref{eq:PE_L2}) by assuming that $P(\tau_{i}|\tau_{i+1})\approx P(\tau_{i}|1)$, because $\tau_{i+1}=1$ will contribute the most. We get
\begin{eqnarray}
  P_{\Delta t}(E)&\propto& F(\Delta t|1)^{E-1},\\
  F(\Delta t|\tau')&\equiv& \sum_{\tau=1}^{\Delta t}P(\tau|\tau'), 
\end{eqnarray}
where $\tau'$ denotes the previous interevent time. This approximation is compared to the numerical results in Fig.~\ref{fig:sequentialML_mu0_1}(e). In addition, for the autocorrelation function we numerically find $\gamma=0.81(1)$ in Fig.~\ref{fig:sequentialML_mu0_1}(f) that fits the scaling relation $\alpha+\gamma=2$ with $\alpha=1.2$ for the regime of large $\tau$ in Eq.~(\ref{eq:alphaL2}). It implies that the dependency between consecutive interevent times is not strong enough for leading to the violations of the scaling relation $\alpha+\gamma=2$.

In general, we have $L$ terms in the memory function:
\begin{equation}
  m(t|\{\tau_i\})= \frac{1}{t}+ \sum_{i=1}^{L-1} \frac{1}{t+s_i},
\end{equation}
where $s_i\equiv \sum_{j=1}^i\tau_j$ with $\tau_j$ denoting the time interval between the $j$th latest and $(j+1)$th latest events. We straightforwardly get
\begin{eqnarray}
  \label{eq:PtauL}
  P(\tau|\{\tau_i\})&\approx& e^{-\mu f(\tau|\{\tau_i\})-\tau/\tau_c} \left[\mu g(\tau|\{\tau_i\})+\tau_c^{-1}\right],\nonumber\\
  \\
  f(\tau|\{\tau_i\})&\equiv &\ln \tau+\sum_{i=1}^{L-1}\ln \left(1+\frac{\tau}{s_i}\right),\\
  g(\tau|\{\tau_i\})&\equiv& \frac{1}{\tau}+ \sum_{i=1}^{L-1} \frac{1}{\tau+s_i}.
\end{eqnarray}
For $s_{k-1}\ll \tau\ll s_k$, we get
\begin{eqnarray}
  f(\tau|\{\tau_i\})&\approx & k\ln \tau - \sum_{i=1}^{k-1}\ln s_i,\\
  g(\tau|\{\tau_i\})&\approx & \frac{k}{\tau}+ \sum_{i=k}^{L-1} \frac{1}{s_i-s_{k-1}},
\end{eqnarray}
leading to $P(\tau)\sim \tau^{-\alpha}$ with $\alpha=1+k\mu$. Similarly to the case of $L=2$, one can infer the scaling behavior of $P(\tau)\sim \tau^{-\alpha}$ as following:
\begin{equation}
  \alpha=\left\{\begin{tabular}{ll}
    $1+\mu$ & for $\tau\ll\tau_{\times 1}$,\\
    $1+2\mu$ & for $\tau_{\times 1}\ll\tau\ll\tau_{\times 2}$,\\
    $\vdots$ & $\vdots$\\
    $1+L\mu$ & for $\tau_{\times L-1}\ll\tau\ll L\mu\tau_c$,\\
    $L\mu$ & for $L\mu\tau_c\ll\tau\ll\tau_c$.
  \end{tabular}\right.
\end{equation}
with crossover interevent times $\tau_{\times i}$ for $i=1,\cdots,L-1$, provided that $\tau_{\times 1}\ll \cdots \ll \tau_{\times L-1}\ll L\mu\tau_c\ll \tau_c$. This implies that the scaling behavior cannot be described by a single value of power-law exponent. We instead calculate the local exponent $\alpha_{\rm local}$,
\begin{equation}
  \alpha_{\rm local}(\tau)=-\frac{\ln P(u\tau)-\ln P(\tau)}{\ln (u\tau)-\ln \tau},
\end{equation}
with a proper constant $u\approx 3.3$. Indeed, such local exponents show gradually increasing behaviors as $\tau$ increases for the cases of large $L$, shown in the insets of Fig.~\ref{fig:sequentialML_mu0_1}(g,j). 

The bursty train distribution can be written as
\begin{eqnarray}
  P_{\Delta t}(E)&=& C'\sum_{\tau_0,\tau_E=\Delta t+1}^\infty \sum_{\tau_1,\cdots,\tau_{E-1}=1}^{\Delta t} \sum_{\tau_{E+1},\cdots,\tau_{E+L-2}=1}^{\infty} \nonumber \\
  & & P(\tau_E) \prod_{i=0}^{E-1} P(\tau_{i}|\tau_{i+1},\cdots,\tau_{i+L-1}), \label{eq:PE_L}
\end{eqnarray}
where $C'$ is a normalization constant and $\tau_{E+1}$, $\cdots$, $\tau_{E+L-2}$ are dummy variables once $\tau_E>\Delta t$. For small $\Delta t$, by assuming that 
$P(\tau_i|\tau_{i+1},\cdots,\tau_{i+L-1})\approx P(\tau_i|1,\cdots,1)$, one gets $P_{\Delta t}(E)$ being proportional to $F(\Delta t|1,\cdots,1)^{E-1}$ with
\begin{equation}
  F(\Delta t|\{\tau'\})\equiv \sum_{\tau=1}^{\Delta t}P(\tau|\{\tau'\}),
\end{equation}
where $\{\tau'\}$ denotes the set of $L-1$ previous interevent times. This result implies the exponential bursty train distribution. For large $\Delta t$, we numerically find scaling behaviors $P_{\Delta t}(E)\sim E^{-\beta}$ with $\beta\approx 1.55(5)$ for $L=10$ and $1.46(3)$ for $L=100$, but limited to the range of $E<L$ as depicted in Fig.~\ref{fig:sequentialML_mu0_1}(h,k). $P_{\Delta t}(E)$ has a natural exponential cutoff $E_c\approx L$ because $L$ directly controls the range of memory effect. The autocorrelation functions for general $L$ also show scaling behaviors with $\gamma\approx 0.58(1)$ for $L=10$ and $0.53(1)$ for $L=100$ in Fig.~\ref{fig:sequentialML_mu0_1}(i,l). Since interevent time distributions are not described by a single value of power-law exponent, and the memory effect induces dependency between interevent times, we do not expect the scaling relation $\alpha+\gamma=2$ to hold.

\begin{figure}[!t]
  \includegraphics[width=.5\columnwidth]{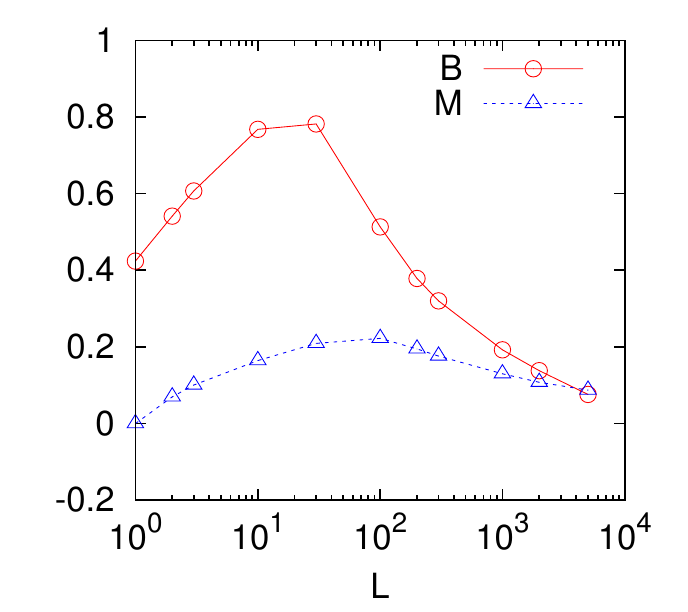}
  \caption{(Color online) Estimated values of burstiness parameter $B$ in Eq.~(\ref{eq:burstiness}) and memory coefficient $M$ in Eq.~(\ref{eq:memoryCoeff}) for different values of $L$ in the model with sequential memory loss mechanism. We used $\mu=1/10$.}
    \label{fig:sequentialML_MB}
\end{figure}

\begin{figure*}[!t]
  \includegraphics[width=.75\textwidth]{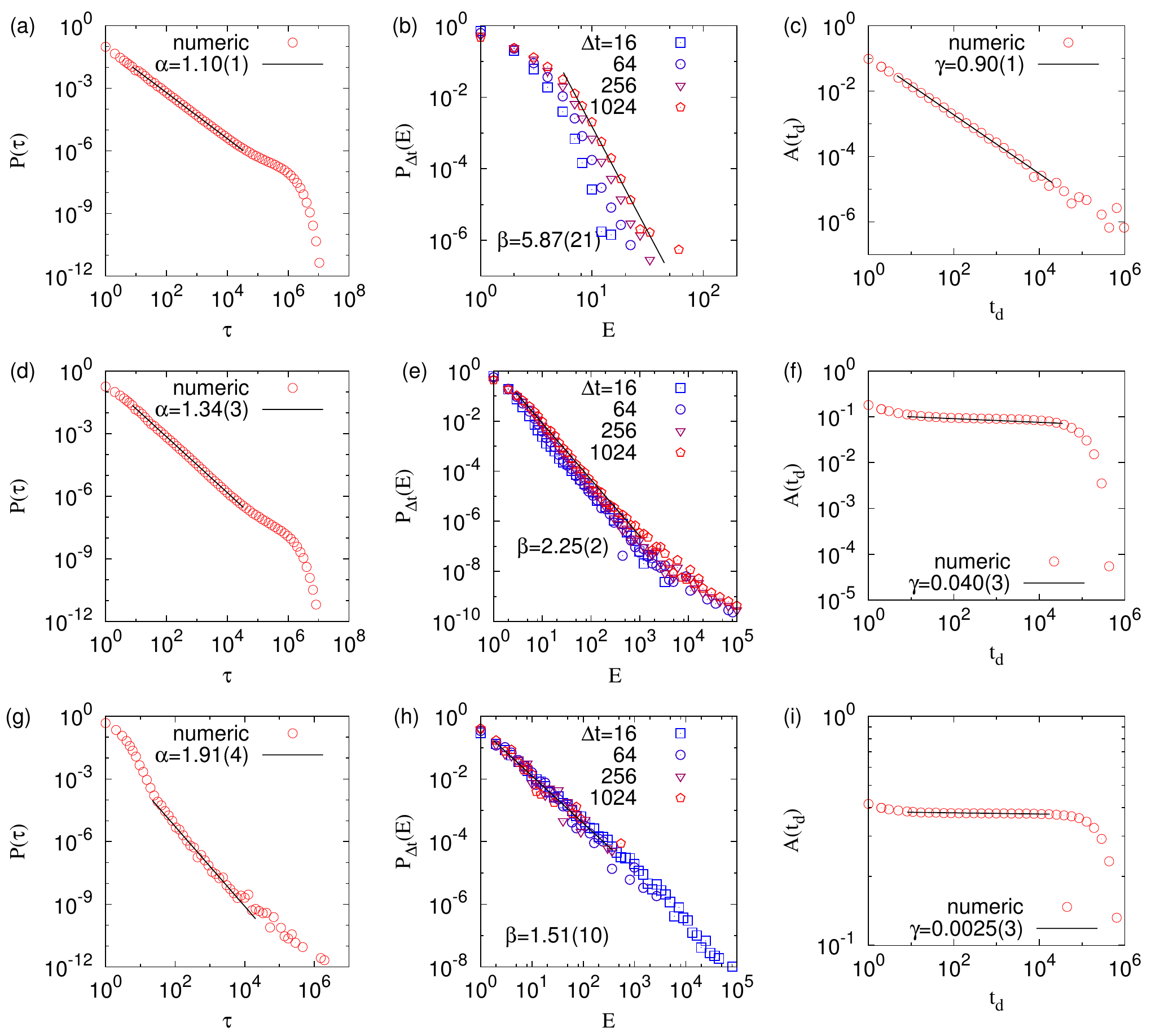}
  \caption{(Color online) Interevent time distributions $P(\tau)$ (left), bursty train distributions $P_{\Delta t}(E)$ (center), autocorrelation functions $A(t_d)$ (right) in the model with preferential memory loss mechanism. We used $\mu=1/10$ and $\nu=3$, $1$, and $0.1$ (from top to bottom). The value of $\beta$ was measured for $\Delta t=1024$ in all cases.}
    \label{fig:preferentialML_summary}
\end{figure*}

\begin{figure*}[!t]
  \includegraphics[width=.75\textwidth]{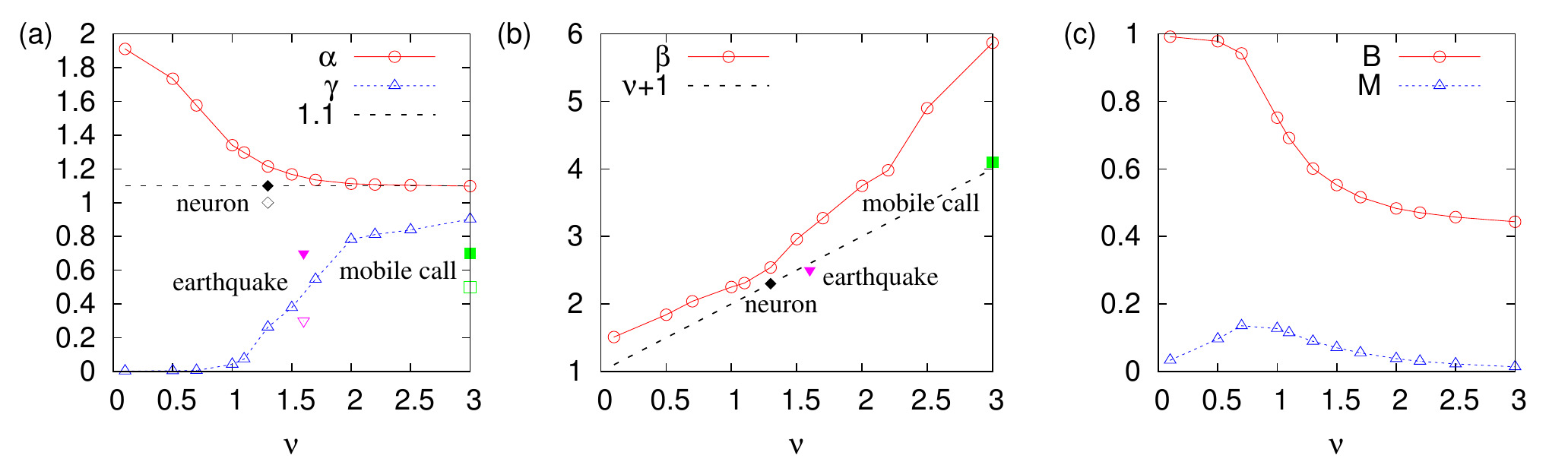}
  \caption{(Color online) Estimated values of $\alpha$, $\beta$, and $\gamma$ (a,b), and $B$ and $M$ (c) for different values of $\nu$ in the model with preferential memory loss mechanism. We used $\mu=1/10$. The value of $\beta$ was measured for $\Delta t=1024$ in all cases. We also plot the empirical values of $\alpha$, $\beta$ (filled symbols), and $\gamma$ (empty symbol) for neuron firings (diamond), earthquakes in Japan (inverse triangle), and mobile calls (square)~\cite{Karsai2012}.}
  \label{fig:preferentialML_expo}
\end{figure*}

\begin{figure}[!t]
  \includegraphics[width=\columnwidth]{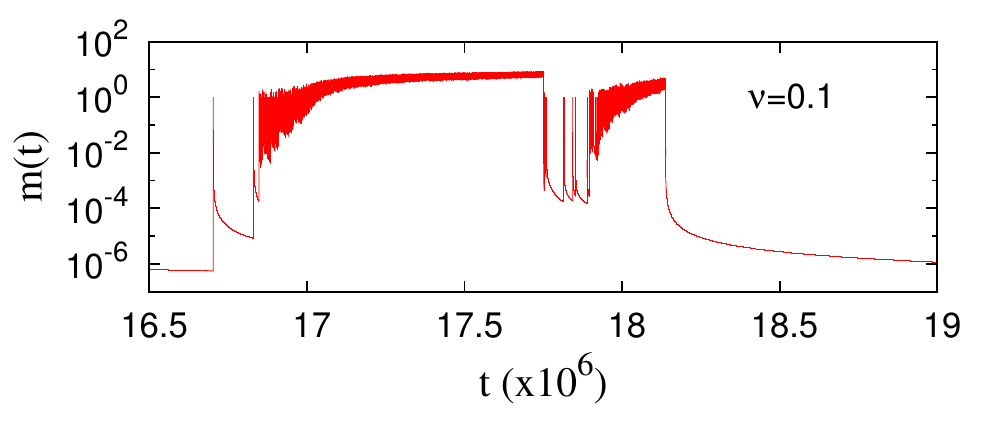}
  \caption{(Color online) Temporal evolution of memory function $m(t)$ in the model with preferential memory loss mechanism. We used $\mu=1/10$ and $\nu=0.1$.}
    \label{fig:preferentialML_memory}
\end{figure}

Finally, let us consider the extreme case of $L\to \infty$. As all the past events contribute to the memory function, the fluctuation of $m(t)$ must be considerably reduced so that the system eventually shows the memoryless Poissonian behavior, as supported by the decreasing fluctuation of $m(t)$ as $L$ increases in Fig.~\ref{fig:sequentialML_memory}. In order to test our expectation, we measure burstiness parameter $B$ in Eq.~(\ref{eq:burstiness}) and memory coefficient $M$ in
Eq.~(\ref{eq:memoryCoeff}) for the wide range of $L$. As depicted in Fig.~\ref{fig:sequentialML_MB}, we find that both $B$ and $M$ increase and then decrease with increasing $L$, implying that there exists an optimal range of $L$ ($30<L<50$) maximizing burstiness and memory effect between interevent times. However, such optimal values of $L$, which play the role of cutoff in bursty train distributions, turn out to be too small compared to the empirical observations, e.g., in~\cite{Karsai2012}.

\subsection{Preferential memory loss}

In order to overcome the strong exponential cutoffs due to $L$ itself, we study the preferential memory loss mechanism using Eq.~(\ref{eq:q_L}). The number of past events contributing to the memory function until a new event occurs is now a random variable, denoted by $L$. The distribution of $L$ is given by $P(L)\propto L^{-\nu-1}e^{-L/L_c}$ with $L_c=\epsilon_L^{-1}$. The interevent time distribution can be obtained from
\begin{equation}
  \label{eq:preferentialPtau}
  P(\tau)=\sum_{L=1}^\infty P_L(\tau)P(L),
\end{equation}
where $P_L(\tau)$ denotes the interevent time distribution for fixed $L$ in the model with sequential memory loss mechanism, i.e., Eq.~(\ref{eq:PtauL}) but with $\{\tau_i\}$ replaced by $\{\tau_{\times i}\}$. Numerical results for $\mu=1/10$ and for several values of $\nu$ are shown in Fig.~\ref{fig:preferentialML_summary}, and the estimated values of $\alpha$, $\beta$, and $\gamma$ as functions of $\nu$ are plotted in Fig.~\ref{fig:preferentialML_expo}(a,b). As $\nu$ increases, $\alpha$
monotonically decreases, while $\beta$ and $\gamma$ monotonically increase. The value of $\beta$ turns out to be larger than that of $\nu+1$. Since $E_c\approx L$ for each $L$, it is expected that $\beta\geq \nu+1$. We also find that the scaling behavior in bursty train distributions is more robust with respect to the value of $\Delta t$, hence comparable to the empirical observations. The empirical values of power-law exponents are plotted in Fig.~\ref{fig:preferentialML_expo}(a,b) for comparison~\cite{Karsai2012}. 

If $\nu$ is sufficiently large, i.e., $\nu\geq 2$, the term for $L=1$ becomes dominant in Eq.~(\ref{eq:preferentialPtau}), leading to $P(\tau)\approx P_1(\tau)\sim \tau^{-\alpha}$ with $\alpha=1+\mu=1.1$ from Eq.~(\ref{eq:alphaL1}). This is consistent with observations that as $\nu$ increases, $\gamma$ approaches $2-\alpha=0.9$ and $\beta$ increases considerably, implying that the bursty train distribution becomes exponential. The values of burstiness parameter $B$ and memory coefficient $M$ get also closer to those for the model with sequential memory loss mechanism with $L=1$. 

For $\nu<2$, as $\nu$ approaches $0$ from $2$, we find that the tail of interevent time distribution becomes thinner because $P(\tau)$ in Eq.~(\ref{eq:preferentialPtau}) is more influenced by the terms of $P_L(\tau)$ with $L>1$, which typically have larger values of power-law exponent. This is evidenced by the increasing behavior of $\alpha$ from $1.1$ to $2$ in Fig.~\ref{fig:preferentialML_expo}(a). Since the very small $\nu$ leads to the very small $q(L)$ in Eq.~(\ref{eq:q_L}), the memory
function is rarely initialized so that some parts of the event sequence can be approximated by the model with the sequential memory loss mechanism for very large $L$. We indeed observe for $\nu=0.1$ that the event sequence is made of dense event clusters spanning relatively long periods separated by long interevent times, as partly depicted in Fig.~\ref{fig:preferentialML_memory}. In each dense event cluster $m(t)$ overall increases, but such non-stationary periods are limited to the timescale of the order of $\epsilon_L^{-1}=10^6$. As $m(t)$ increases but very slowly, it can be roughly approximated by a Poissonian process, supported by the estimation of $\gamma\approx 0$ and $M\approx 0$ in Fig.~\ref{fig:preferentialML_expo}. In addition to $\gamma\approx 0$, the autocorrelation function remains finite for wide range of $t_d$ [see Fig.~\ref{fig:preferentialML_summary}(i)] because of the non-stationarity in $m(t)$ in dense event clusters. Note that the scaling relation $\alpha+\gamma=2$ seems to hold for $\nu=0.1$ even when $\beta\approx 1.5$ and $B\approx 1$, implying strong dependency between interevent times and strong burstiness effect. These can be understood as follows. As bursty trains are mostly measured in dense event clusters, they tend to contain more events, leading to the heavier tail for bursty train distributions and the smaller value of $\beta$, i.e., $\beta\approx 1.5$. Relatively few but very large interevent times separating dense event clusters force $B$ to get close to $1$. 

\section{Conclusions}

In order to investigate the underlying mechanism behind correlated bursts widely observed in natural phenomena and human activities, we have devised and studied a simple model that is able to generate correlated bursts using a self-exciting point process with variable range of memory. Our model does not need to declare the bursty trains as compared to the previous two-state model for correlated bursts. In our model, a new event can occur depending on the memory function defined as the sum of decaying memories of past events. For incorporating noise and/or the limited memory capacity of systems, we apply two different memory loss mechanisms: fixed number or variable number of memories, which we call sequential and preferential memory loss mechanisms, respectively. For each case, we obtain the interevent time distribution, bursty train distribution, and autocorrelation function, all of which are characterized by power-law decaying exponents $\alpha$, $\beta$, and $\gamma$, respectively, to study scaling relations among them.

For the model with sequential memory loss mechanism, the memory function is given by the sum of decaying memories of $L$ latest events, where $L$ is a control parameter. The simplest case with $L=1$ has been exactly solved, also satisfying the scaling relation $\alpha+\gamma=2$~\cite{Vajna2013}. Other simple cases could be analytically solved, while the general cases have been numerically studied. As $L$ becomes larger, the bursty train distribution shows scaling behavior for limited range of parameters, implying the emergence of correlated bursts. However, the number of events in bursty trains is strongly limited by $L$. Interestingly, if $L$ is extremely large, too much memory effect effectively reduces the model to the Poisson process, which is confirmed by both memory coefficient and burstiness parameter approaching $0$, i.e., no memory effect and no burstiness. 

In order to overcome the strong cutoff effect due to the fixed $L$, we have numerically studied the model with preferential memory loss mechanism. Here the number of memories $L$ in the memory function increases gradually but is set as $1$, i.e., memory function initialization, with probability controlled by the exponent $\nu$. For sufficiently large $\nu$, the memory function is initialized frequently so that the model can reduce to the case with sequential memory loss mechanism using $L=1$. On the other hand, for very small $\nu$, the event sequence is composed of dense event clusters spanning long periods that are separated by very large interevent times. Dense event clusters may correspond to the case with sequential memory loss mechanism using very large $L$, i.e., close to the Poisson process. For the intermediate range of $\nu$, we find evidences that our model generates correlated bursts, hence comparable to the empirical findings.

As a followup, our models can be extended to incorporate a number of complex realistic situations. For example, we can consider the context of events~\cite{Jo2013}, and a network of interacting individuals, each of which shows activities with correlated bursts.

\begin{acknowledgments}
HJ gratefully acknowledges financial support by Aalto University postdoctoral program and by Mid-career Researcher Program through the National Research Foundation of Korea (NRF) grant funded by the Ministry of Science, ICT and Future Planning (2014030018), and Basic Science Research Program through the National Research Foundation of Korea (NRF) grant funded by the Ministry of Science, ICT and Future Planning (2014046922). JIP acknowledges support by the Academy of Finland, project No. 260427. JK acknowledges support from H2020 FETPROACT-1-2014, grant agreement \#641191 ``CIMPLEX''.
\end{acknowledgments}


\end{document}